\documentclass[manuscript,screen]{acmart}
\usepackage{xcolor}
\usepackage{soul}
\usepackage{balance}
\usepackage{graphicx}

\AtBeginDocument{%
  \providecommand\BibTeX{{%
    \normalfont B\kern-0.5em{\scshape i\kern-0.25em b}\kern-0.8em\TeX}}}

\setcopyright{acmcopyright}
\copyrightyear{2018}
\acmYear{2018}
\acmDOI{XXXXXXX.XXXXXXX}

\acmConference[Conference acronym 'XX]{Make sure to enter the correct
  conference title from your rights confirmation emai}{June 03--05,
  2018}{Woodstock, NY}
\acmPrice{15.00}
\acmISBN{978-1-4503-XXXX-X/18/06}

\begin{document}

\title{"I Upload...All Types of Different Things to Say, the World of Blindness Is More Than What They Think It Is": A Study of Blind TikTokers' Identity Work from a Flourishing Perspective}

\author{Yao Lyu}
\orcid{0000-0003-3962-4868}
\affiliation{%
  \institution{Pennsylvania State University}
  \city{University Park}
  \state{Pennsylvania}
  \country{USA}
}
\email{yaolyu@psu.edu}

\author{Jie Cai}
\orcid{0000-0002-0582-555X}
\affiliation{%
  \institution{Pennsylvania State University}
  \city{University Park}
  \state{Pennsylvania}
  \country{USA}
}
\email{jpc6982@psu.edu}

\author{Bryan Dosono}
\orcid{0000-0002-1312-1034}
\affiliation{%
  \institution{eBay, Inc.}
    \city{New York City}
  \state{New York}
  \country{USA}
}
\email{bdosono@ebay.com}

\author{Davis Yadav}
\affiliation{%
  \institution{Pennsylvania State University}
  \city{University Park}
  \state{Pennsylvania}
  \country{USA}
}
\email{kzl5736@psu.edu}

\author{John M. Carroll}
\orcid{0000-0001-5189-337X}
\affiliation{%
  \institution{Pennsylvania State University}
  \city{University Park}
  \state{Pennsylvania}
  \country{USA}
}
\email{jmcarroll@psu.edu}

\renewcommand{\shortauthors}{}


\begin{abstract}

Identity work in Human-Computer Interaction (HCI) has focused on the marginalized group to explore designs to support their asset (what they have). However, little has been explored specifically on the identity work of people with disabilities, specifically, visual impairments. In this study, we interviewed 45 \textit{BlindTokers} (blind users on TikTok) from various backgrounds to understand their identity work from positive design perspective. We found that BlindTokers leverage the affordance of the platform to create positive content, share their identities, and build the community with the desire to flourish. We proposed \textit{flourishing labor} to present the work conducted by BlindTokers for their community's flourishing with implications to support the flourishing labor. This work contributes to understanding blind users' experience in short video platforms and highlights that flourishing is not just an activity for any single Blind user but also a job that needs all stakeholders, including all user groups and the TikTok platform, serious and committed contribution. 
    \end{abstract}




\maketitle

\section{Introduction}

The last decades have witnessed the rapid development of various social media platforms. Today, using social media to share personal aspects of life has become a part of the daily routine of many social media users. The Human-Computer Interaction (HCI) research community has been paying constant attention to users' presentation of everyday lives on social media platforms \cite{simpson_for_2021,simpson_how_2022, 10.1145/1978942.1979381,Scheuerman2020}. HCI scholars believe that such presentations reflect users' attitudes on their conception of themselves, and the activities can also be mediated by platforms' technical affordance. The HCI literature has especially documented a research stream that focuses on social media users' collective activities at the community level in which the activities reveal various aspects of a specific group of users' shared characteristics. One of the most important lenses is collective identity, as it represents how people conceptualize themselves on social media and connect themselves to each other. Collective identity is "\textit{a person’s sense of belonging with a group; where being part of that group is part of how an individual sees themselves} (\cite{dasCollaborativeIdentityDecolonization2022}, p.3)". The work involved in presenting such identity by the user group is called identity work. 

Recently, the Computer-Supported Cooperative Work (CSCW) and HCI literature show a growing interest in the identity work by minority groups, like LGBTQ+ \cite{dym_coming_2019,simpson_for_2021,devito_too_2018}, Black people \cite{10.1145/1978942.1979381}, and Asian and Pacific Islander (AAPI) communities \cite{zhang_chinese_2022,dosono_aapi_2018,dosono_exploring_2017,dosono_moderation_2019,Dosono2018}. Critical identity work represents the user groups that are oftentimes marginalized by mainstream society. The identity work helps these groups to voice out and counter marginalization. Especially, there is growing literature specifically on people with disabilities \cite{kameswaran_advocacy_2023,heung_nothing_2022}. The population of people living with disabilities is significant. As an example, people with visual impairments (PVI) number around 10 million US adults according to the CDC \cite{cdc_disability_2022}. In addition, many blind people are active on social media platforms. According to the authors' search in late 2022, TikTok surged in usage as one of the most popular social media platforms in the world and amassed approximately 5 million videos tagged with "blind" or other variations related to people with visual impairments. The vast user base on TikTok helps blind users form a community specifically for themselves called "BlindTok\footnote{To be noted, we use "people with visual impairments" as the term to describe the population; we also use BlindTokers or blind TikToker, since the community of study references the term "BlindToker" or "blind TikTokers" to describe themselves on TikTok.}." The blind users are named "BlindTokers." The large community of blind users on a visual platform like TikTok warrants an investigation of how they conceptualize and construct their identities on social media, especially at the community level. Therefore, we deploy a study focusing on the collective identity work of blind users on TikTok with a central research question:

\begin{center}\textbf{\textit{RQ: What was the collective identity work of BlindTokers?}}\end{center}

The current research is part of a long-term ethnographic project that examines blind users' experiences with TikTok. In total, we interviewed 45 blind TikTok users who came from a wide variety of backgrounds. We deployed thematic analysis \cite{Braun2006b} to study the interviewees' experiences. After the analysis, we reported three themes: demonstrating BlindTokers' characteristics, voicing out BlindTokers' rights and responsibilities, and addressing sighted people's misconceptions. The first theme shows how users introduced their characteristics to the public as blind people; the second theme presents participants' work on deliberating their rights and responsibilities as blind users of TikTok; the last theme demonstrates the tension between blind and sighted communities and how BlindTokers addressed it. 



Our contributions to the CSCW and HCI literature are multi-faceted. First, we present three themes of BlindTokers' identity work with detailed descriptions and TikTok's role in such work. Second, we used flourishing in the context of positive design as a lens to underscore BlindTokers' desires for positivity. We also applied a labor perspective (coined "flourishing labor") to unpack BlindTokers' effort to support the state of flourishing. This study supplements recent work about positive design with marginalized groups \cite{to_flourishing_2023} and contributes to a more nuanced understanding of blind users' identity on short-video platforms.

\section{Interfaces of TikTok}
\begin{figure*}[htp]
    \centering
    \includegraphics[width=\textwidth]{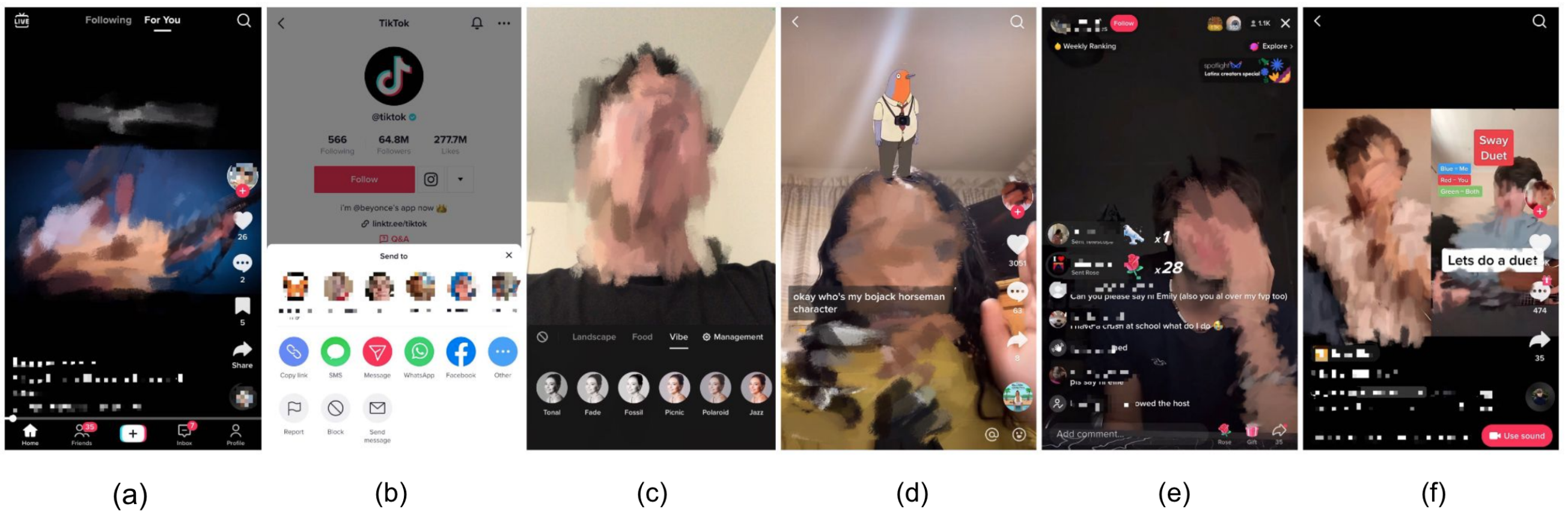}
    \caption{TikTok's Interfaces, including (a) Primary Interface, (b) User Profile, (c) Video Filter, (d) Character Match, (e) Live-stream, and (f) Duet.}
    \label{tiktok}
    \Description{TikTok's Interfaces}
\end{figure*}

In this section, we briefly introduce TikTok's interfaces relevant to BlindTokers' experiences. The primary interface of TikTok includes a video that almost occupies the entire screen; buttons, such as the video creator's profile picture, "like," "comment," and "share," on the left side; the titles, descriptions, and hashtags of the video at the bottom; and "following," "for you" at the top (Figure \ref{tiktok}, a). To view one's homepage, users can click on their profile picture or names. The homepage presents their username, their bio, and their videos published (Figure \ref{tiktok}, b). Viewers can share the page through links, SMS, or direct messages to other TikTokers. Creators can select and pin certain videos at the top of the page. At the bottom of the primary interface, users can tap the "+" button to start creating videos (Figure \ref{tiktok}, a). On the video creation page, users can use the phone camera to take videos and add visual and/or audio effects (Figure \ref{tiktok}, c).


In addition to adding effects, TikTok also has built-in functions that make creating videos more playful. For instance, users can create videos of their faces and then let the algorithm match the faces with characters from cartoons, movies, etc. (Figure \ref{tiktok}, d). To enhance the viewer experience, creators can also add captions to their videos (Figure \ref{tiktok}, f). TikTok also allows users to interact with others through various functions such as live-streaming, stitching, and duets. The live streaming interface is similar to the interface for displaying videos. The main difference is that it displays the commenting and gifting behaviors of viewers in real time, so the lower part of the interface automatically scrolls to present the most recent comments or gifts (Figure \ref{tiktok}, e). Duet is a type of activity that requires users to create videos collaboratively. One user creates a video in advance, and another user "duets" it by putting the video on the right side of the screen and creating another according to the previous one. Usually, duets are manifested as co-performance like singing or dancing together. For collaborative singing, sometimes singers who create videos for people to duet later use different colors to mark the lyrics so that later creators know which part they need to sing (Figure \ref{tiktok}, f).


\section{Related Work}

\subsection{Blind User and Online Community}
The development of ICTs has made it possible for blind people to access social media websites or applications where the content is mostly visual \cite{prajwal_towards_2019,lee_revisiting_2019,gleason_addressing_2019,gleason_twitter_2020}. Blind users can use screen readers to read visual information, such as texts or images  \cite{barbosa_every_2022,jain_smartphone_2021,abdolrahmani_should_2016}. That said, screen readers cannot resolve all accessibility issues because of blind users' situated requirements on visual information. Stangl et al. \cite{stangl_person_2020} introduced an investigation on blind people's specific needs in terms of understanding images in different contexts: when users were viewing images posted on general social network sites, they cared about who was in the picture and what they were doing; while for pictures on dating sites, users were interested in what the people in the pictures looked like, including the hairstyle, body figure, and whether people met certain beauty standards.

The situated requirements when using social media further point out that blind users' experiences are technically supported and socially constructed. The experiences are entangled with not only the accessibility designs but also specific social contexts, such as playing games \cite{stadler_blind_2018}, hosting live-streaming \cite{jun_exploring_2021,Li2022}, understanding emojis \cite{tigwell_emoji_2020} and memes \cite{gleason_making_2019} in communication, etc. With such contextual requirements, blind users often seek help from others, especially from other blind users \cite{brady_investigating_2013}. Such interactions cultivate online communities for blind users, and existing research articulates how experiences impact the online community of blind users \cite{seo_exploring_2017,seo_understanding_2021}. Rong et al. \cite{rong_it_2022} reported a study on blind users on Douyin, the Chinese version of TikTok. Blind users experienced marginalization by Douyin's algorithm: they connected together to form a community, and they then provided and received support in the community. The research on the community of blind users presents a group of people with similar ability statuses. They have similar characteristics and lifestyles.

The prevalence of social media platforms has made online activities a vital part of most users' everyday routines. Therefore, a growing number of studies argue that social media platforms should be considered the infrastructure of online users' daily lives \cite{Zhou2020,Plantin2018,Plantin2019}. In the online space, which is supported by social media platforms, users "live" with their communities; they conduct various social activities with other users to fulfill their lives. In this study, we report a case where blind users of TikTok form a community, called BlindTok, specifically for themselves. We present how the users conducted various activities in the community. We report how BlindTokers conceptualize their collective identity.




\subsection{Identity and Flourishing}


We also analyze BlindTokers' social interactions from an identity perspective, especially the positive side of identity. Collective identity is "a person’s sense of belonging with a group; where being part of that group is part of how an individual sees themselves (\cite{dasCollaborativeIdentityDecolonization2022}, p.3)." The HCI community has been constantly investigating how users perceive and construct their identities. The identity includes various aspects, such as race \cite{to_they_2020,smith_whats_2020}, culture \cite{10.1145/3579509}, religion \cite{salehi_sustained_2023}, and gender \cite{simpson_for_2021,simpson_how_2022,das_jol_2021}. HCI scholars shed light on how groups historically excluded from mainstream society construct their identity on social media platforms \cite{das_decolonial_2022,dosono_aapi_2018}. Most of research expounds a damage-centered approach, which critically examines the current status of targeted groups identity with an emphasis on what they are lacking \cite{cunningham_grounds_2023,yosso__whose_2005}; recently, a strand of research \cite{to_flourishing_2023,klassen_more_2021,10.1145/1357054.1357225} has shown increasing interest in the positive side of identity, for instance, the assets of a community. To better illustrate the theoretical foundation of the positive side of identity, we also turn to the literature on the notion flourishing in positive design.


Positive design is a design framework for subjective well-being. It focuses on the positive aspects of human life, such as happiness. In "Positive Design: An Introduction to Design for Subjective Well-Being \cite{desmet_introduction_2013}," Desmet and Pohlmeyer provided a framework for positive design that consists of three ingredients: design for pleasure, personal significance, and virtue. \textit{Pleasure} refers to happiness based on enjoyment, which could be operationalized as the increase of positive sentiments and/or the decrease of negative sentiments. And usually, pleasure is momentary (short-term). \textit{Personal significance} emphasizes the individual meaning side of happiness. It can be obtained through a sense of making progress that drives the individual toward their desired achievements. Therefore, personal significance can be short-term or long-term, depending on which type of goals (short-term or long-term) the individual wants to achieve. \textit{Virtue} foregrounds the happiness resulting from one's virtuous practices. In other words, people would feel happy when their behavior aligns with their morality and ethics. 

\begin{figure*}[htp]
    \centering
    \includegraphics[width=\textwidth]{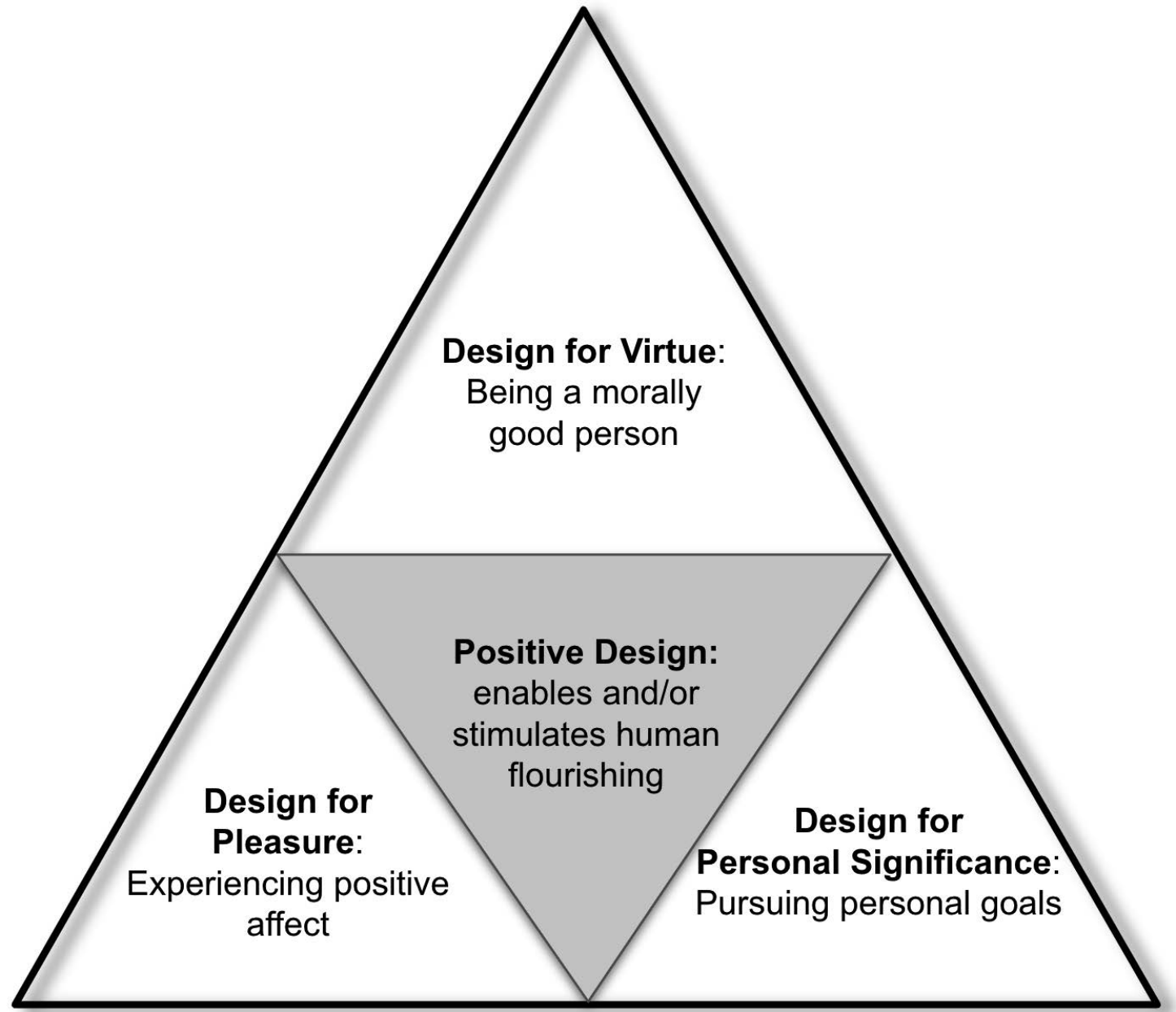}
    \caption{Positive Design Framework (\cite{desmet_positive_2013}, p.5)}
    \label{fig:positive}
    \Description{Positive Design Framework}
\end{figure*}

Further, Desmet and Pohlmeyer pointed out a well-being state which covers all the three ingredients mentioned above, and the state is called \textit{flourishing} (Figure \ref{fig:positive}). Flourishing means one can live to their potential \cite{ryan_happiness_2001}; it "encompasses (self-focused) personal development as well as (other-focused) virtuous living in the best interests of society (p.10) \cite{desmet_introduction_2013}." Recently, To el at. \cite{to_flourishing_2023} added to the flourishing literature by introducing three case studies of smartphone apps that contributed to the flourishing of the Black, Indigenous, and People of Color (BIPOC) community. They specifically pointed out the socio-technical affordance of such apps and the values produced by users with the support of such affordance. For instance, the study introduced an app called \textit{The Wave}, which focused on cultivating an environment for adult Black communities to share their everyday lived experiences. The app's features, such as invitation-only registration methods and shared calendars, cultivated a value of cultural wayfinding, i.e., to "promote convenience, safety, and support for communities that may not experience these constructs in other spaces (p.923) \cite{to_flourishing_2023}." To et al. \cite{to_flourishing_2023} also proposed several tenets for designing for flourishing, including designing for self-actualization, desire to flourish, and self-sustainability. 

In this paper, we consider flourishing as a specific aspect of positive experience that BlindTokers desired. We analyze BlindTokers' conception of their collective identity from a flourishing perspective. Especially, by highlighting the positivity in BlindTokers' identity work, we turn away from the frameworks that focus on the deficits of the community of people with disabilities. In stead, we embrace a framework that focuses on how BlindTokers cherish, enjoy, and support as a community.

\subsection{Identity Work From A Labor Perspective}
Finally, we also review relevant research on identity work and labor. Research on identity has pointed out that constructing one's identity requires intentional and organized practices. The practices are called "\textit{identity work}." Snow and David defined identity work as "the range of activities individuals engage in to create, present, and sustain personal identities that are congruent with and supportive of the self concept (p.1348) \cite{snow_identity_1987}." Identity work contributes to "forming, repairing, maintaining, and strengthening or revising their identities to preserve existing identities and/or externally ascribed self concepts (p.849) \cite{morioka_identity_2016}." Snow and David further introduced four types of identity work: 1) arranging physical items or settings to fit the identity; b) managing personal appearances; 3) selecting and joining particular communities; and 4) verbal construction and assertion of personal identities.

Identity work, despite its variety, involves many concerns \cite{alharbi_accessibility_2023,rovira-sancho_activism_2023}. For instance, in analyzing self-presentation of identity, Erving Goffman \cite{goffman_presentation_2016} distinguished the front-stage behaviors and back-stage behaviors to illustrate the invisible part (back-stage) of identity work. The invisible identity work could be less recognized and credited, hindering sustainability of identity work \cite{to_flourishing_2023}. Reflecting on this concern, some scholars proposed a labor perspective to analyze identity work \cite{duffy_nested_2021,scolere_constructing_2018,woodcock_affective_2019,duguay_running_2019}. Labor is the "exertion of the body or mind... usually used to describe activities that have some sort of compulsion attached to them (p.276) \cite{hesmondhalgh_user-generated_2010}." Many scholars in HCI have investigated labor in terms of using ICTs, such as creative labor \cite{simpson_rethinking_2023}, algorithmic labor \cite{ma_how_2021}, access labor \cite{alharbi_accessibility_2023}. Dosono and Semaan \cite{dosono_moderation_2019} reported a study on Asian American and Pacific Islander (AAPI) content moderators on Reddit. The study investigated AAPI's identity work, including redefining public identity, challenging stereotypes, and negotiating diverse identities. The authors pointed out that such identity work could involve adversity and conflict and needed an environment to support sustainability. They proposed an emotional labor perspective to analyze AAPI moderators' practices that cultivated a supportive online space for AAPI's identity work. The study drew attention to moderators' invisible work in terms of managing their own emotions and providing emotional support to peers when moderating the content of AAPI subreddits. By applying a labor perspective, this study unpacked the invisible work involved behind the scenes to sustain online communities. The perspective foregrounded moderators' investment driven by their desire to contribute to the collective good and argued for more awareness of recognizing and rewarding such investment so that moderator contributions can be sustainable. 

We also build our research on the literature at the intersection of identity work and labor. Specifically, we aim to investigate the investment by BlindTokers who worked on supporting the BlindTok community to flourish. By doing so, we want to raise awareness of the invisible labor conducted by such contributors and argue for a better collaboration mode to make the flourishing more sustainable. 






\section{Methods}



\subsection{Interview Protocol}
The current study is part of an ongoing long-term ethnography on the experiences of blind users with TikTok and was approved by our Institutional Review Board (IRB). We used semi-structured interviews to collect data. In the interview protocol, we included questions relevant to 1) social interactions, including interactions among blind users and interactions between blind users and sighted users (e.g., "What content did you post on TikTok?" "What comments did you get from your audience?" "Which content creators did you follow on TikTok?" "How did you interact with the creators?"), 2) the role of TikTok technical affordance (design, functions, features, etc.) in such social interactions (e.g., "How did you use TikTok's recording/stitching/dueting/live-streaming?" "How did you use TikTok's hashtag/comment/like"?), and 3) the challenges BlindTokers encountered in the social interactions and reactions to them (e.g. "What accessibility issues did you encounter when using TikTok?" "How did you address them?").

We focused on the participants' experiences, interactions, and engagement with other TikTok users during the interview. We also paid attention to information related to accessibility issues and solution-seeking. That said, we did not let the protocol confine the interview. We used follow-up questions to explore the experiences of study participants on TikTok. When participants recalled memories or made comments that were not covered in the protocol, we encouraged them to talk more for further probing. For example, when asked about social interactions with other TikTokers, all participants mentioned their interactions with other blind users; this further helped us know about the existence of the \textit{BlindTok} community, which was not covered in the initial interview protocol. We then ask specific questions about the BlindTok community, including 1) perceptions of BlindTok (e.g., "How did you know the BlindTok community?"), 2) interactions within BlindTok (e.g., "What did you do with other BlindTokers?"), and 3) interactions beyond BlindTok (e.g., "As a BlindToker, what were the experiences with sighted TikTokers?").


We conducted all interviews with BlindTokers remotely (either via Zoom or phone calls). Before starting each interview, we informed the participants about the purpose of the study, introduced the research members, and obtained verbal informed consent from each of the participants. The interviews were audio-recorded and then meticulously transcribed into text files. We present verbatim quotes from our study participants in Section \ref{sec:findings}; we also provide additional contextual information to improve the coherence of the text.

\subsection{Sampling}
We recruited all participants in our study from TikTok. We created an official TikTok account for the research team with a real name, affiliations, and information about the current study. Then we searched for visually impaired users' profiles using keywords like "visual impairment," "blind," etc. After identifying blind TikTok users, we followed their accounts and engaged with their content by commenting. During that, we introduced the study objectives and the study team. In addition, we also supported their content creation by liking their videos and attending their live streams. For each BlindToker who followed our account back, we shared an official call for participation (via direct message) in the study. Finally, we used the snowball sampling method to reach more BlindTokers.
 
Through this procedure, we recruited 45 English-speaking BlindTokers. The participants consisted of 24 self-reported females, 18 males, and 3 non-binary individuals. The ages of our participants ranged from 19 to 59 years (an average of 34 years). In terms of the severity of the visual impairment, 35 participants were legally blind, 7 had low vision, and 3 participants were completely blind. Regarding their daily usage of TikTok, 26 participants spent more than 60 minutes, 13 between 30 and 60 minutes, and three spent less than 30 minutes. To protect participants' privacy, we only provide a general report on participants' demographics rather than listing all participants' information in a table.

\subsection{Data Analysis}
In this study, we used thematic analysis \cite{Braun2006b} as an approach to interpreting the data. After documenting the interview data, we thoroughly read the transcripts to get a first impression of the participants' experiences. As we immersed ourselves in the experiences of the participants, we noticed the close relationship between the participants and the BlindTok community. According to the participants, BlindTok was an umbrella term for the overall blind user group on TikTok. Blind TikTok users used various TikTok features to look for and interact with other blind users. The TikTok-mediated social interactions helped blind users form the BlindTok community. Therefore, we paid close attention to the social interactions based on the BlindTok community when we coded the data. The codes were then grouped to form higher-level themes. 

While we were refining the themes, we observed that the interactions around the BlindTok community were closely relevant to the presentation of identity, such as showing what a blind person's daily life was like. Furthermore, we found that the identity work was primarily affected by TikTok functions. Some functions supported identity work and some did not, depending on specific contexts. Finally, we generated three themes. The first theme shows how users introduced their characteristics to the public as blind people; the second theme presents participants' work on deliberating their rights and responsibilities as blind users of TikTok; the last theme demonstrates the tension between blind and sighted communities and how BlindTokers addressed it. More details of the findings and discussion are presented in later sections.

\subsection{Reflexivity Statement}
We also think it is necessary to provide our reflexive statement to acknowledge the potential concerns of the study and to show the work we conducted to minimize the concerns. All researchers in the research team are sighted individuals. The first author, an indigenous Asian male, took a lead role in the project. He has been a volunteer in a local blind community for two years. The last author, a white male researcher who had conducted more than ten projects on improving ICT accessibility for blind users, advised the first author in this project. Other authors also have research experiences relevant to the identity of minority groups. That said, there could be potential bias (e.g. failed to capture some nuanced identity work during coding) due to the lack of blind researchers in our team.

To counter this, we developed a protocol and strictly followed it. 1) To make the content accessible, we used TikTok videos instead of printed materials or emails to promote our project. 2) We reach out to participants based on their self-disclosures. 3) Instead of asking participants to fill out forms, we confirmed the visual impairment status and collected demographic data with them through conversations before the interviews. 4) We carefully chose the language used in the interviews and write-up of the findings; we scheduled phone calls with participants and read the findings to them; we made revisions when participants felt uncomfortable about some phrases. However, we still acknowledge the potential limits of our study.
\section{Findings}\label{sec:findings}

\begin{figure*}[htp]
    \centering
    \includegraphics[width=\textwidth]{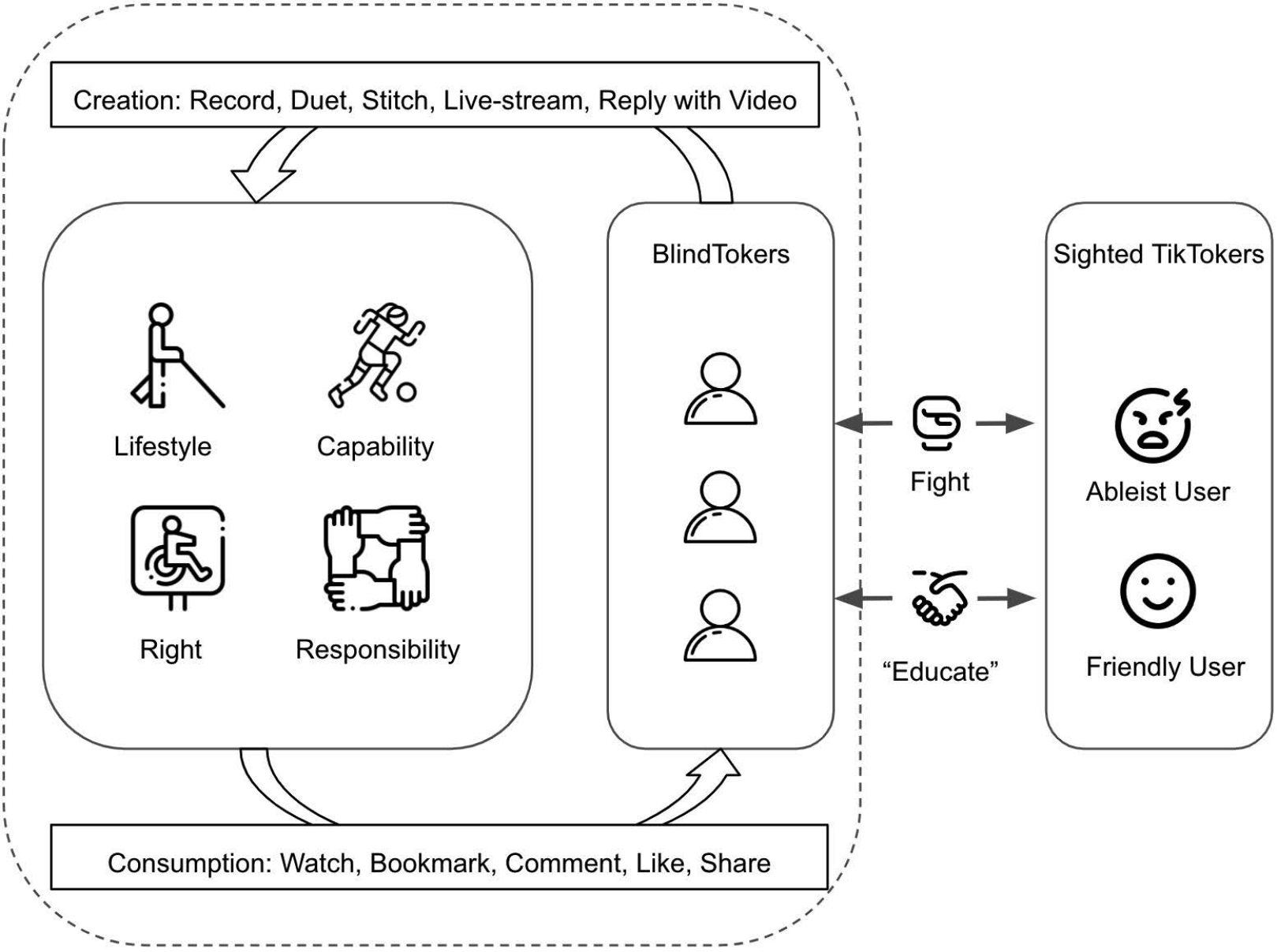}
    \caption{Identity Work of BlindTokers}
    \label{findings}
    \Description{Identity Work of BlindTokers}
\end{figure*}

This section presents three themes to illustrate how BlindTokers conducted identity work (Figure: \ref{findings}). The first theme showcases how participants used TikTok to present their characteristics. The second theme illustrates the participants' efforts to demonstrate their rights and responsibilities as BlindTokers. Finally, we also describe how BlindTokers addressed misconceptions about blind people held by sighted individuals. The report includes not only BlindTokers' content creation and social interactions but also their strategies, attitudes, and feelings while they conducted this identity work. We also illustrate how they utilized TikTok's specific designs during the process.

\subsection{Demonstrating BlindTokers' Characteristics}



For BlindTokers, the most common scenario of using TikTok was self-presentation. Participants utilized TikTok to record and depict their daily lives. BlindTokers enjoyed showcasing their experiences of being blind, encompassing not only various levels of vision but also their individual lifestyles. These characteristics highlighted the distinctive life experiences of blind individuals and further emphasized their unique identity.


\subsubsection{\textbf{Presenting Blind People's Lifestyles}} 


All participants (N=45) reflected on the self-presentation aspect of TikTok. One example was that most participants incorporated terms such as "blind," "cane," "visually-impaired," or related phrases into their usernames to disclose their identity. Regarding content creation, some of them used TikTok to generate content discussing items or events that held special significance within the blind community. Blind individuals sometimes encounter building designs that are inaccessible to them. Therefore, creating videos about such experiences helped the blind community collectively voice their grievances. For instance, P18 created a parody video outlining his encounters with wet floor signs in his videos:

\begin{quote}\textit{[Wet floor sign] is the enemy of people with visual impairments...It doesn't contrast enough. So if we're not looking down...you go in and everyone's there and it makes a really loud bang. When I discovered it wasn't just me who had a problem with them, I make videos on TikTok to make jokes... there's a specific sound I did for one of the most recent videos, like I bump into [the sigh], that's a sound from a movie... mostly people say, `Hey, I totally relate to this.` [P18]}\end{quote}

P18 was a radio host who conducted a series of shows specifically designed for blind people. P18's content creation demonstrated how he, as a blind person, reacted to wet floor signs. While wet floor signs might serve a purpose for sighted individuals, they posed significant risks, embarrassment, and frustration for blind people. The content about wet floor signs further connected him with other blind individuals who had similar experiences. They shared their feelings, providing emotional relief. This also indicated that, instead of being upset by complaints about wet floor signs, blind people found enjoyment in the positive approach of addressing the issue through humor.

Additionally, P18 mentioned that he utilized various TikTok audio effects, such as voice-over acting and special audio effects, to make his content more engaging \cite{schaadhardt_laughing_2023}. This ensured that his content remained accessible to blind people. While some participants created content individually and shared it with others, as P18 did, others collaboratively produced content to present themselves to their followers. For instance, P12, a member of a group of male BlindTokers, shared how he socialized with other group members:

\begin{quote}\textit{We did a couple of duets together, We did dancing, movie scenes. We stitched each other videos asking questions like, "What's your favorite color?" "[What are] similarities other than blindness?" ...I feel accomplished [P12]}\end{quote}

P12 talked about how he hung out with his blind friends on TikTok. They used TikTok's affordances to enjoy communication as a part of the blind community, as well as for exploring more about themselves beyond their blindness. In fact, many blind users consider TikTok a platform for communicating with other blind users worldwide because there are far more blind users on TikTok than on other social media platforms of the era ([P14]). With such a significant number of blind users, BlindTokers not only showcased their lives but also witnessed the lives of other blind people on TikTok. Witnessing others' lives not only fostered a sense of connection (e.g., as seen in the story of P18) but also provided resources for improving self-presentation. For example, P13 recalled the experience of watching and relating to videos made by another blind user to explain vision loss to sighted people:


\begin{quote}\textit{...he (a BlindToker) did the special effect that shows what it looks like to lose your central vision. (When I try to describe my vision)... I showed it, and she was just fascinated and she was like, "I always thought it was just a black hole." And I'm like, no. It's like pixelated, like your brain just creates something there that's made up. [P13]}\end{quote}

Visual impairments manifest in various ways, meaning that people with visual impairments may experience different types of vision or vision loss. BlindTokers utilized TikTok's special effects in their videos to vividly illustrate their perspectives. Additionally, these TikTok videos offered other BlindTokers, who shared similar experiences, valuable materials for enhancing the portrayal of their lives.

\subsubsection{\textbf{Showcasing Capabilities as BlindTokers}} 


In addition to describing their characteristics, more than half of the participants (N=32) had either engaged in showcasing their abilities through content creation on TikTok or in consuming content featuring capability demonstrations by BlindTokers. According to the participants, demonstrating their capabilities was an integral part of their identity presentation. This not only conveyed their aspirations for independent living but also served as inspiration for other blind users. Consequently, some participants regarded the presentation of their abilities as a means to represent themselves and to foster a shared identity among a community of BlindTokers. For instance, P03 shared how she used TikTok videos to express her understanding of being blind:

\begin{quote}\textit{Sometimes I upload a video on the topic of my blindness. Sometimes I upload videos of me cooking, all types of different things, to say like, the world of blindness is more than what they think it is, that we have regular lives. So I've uploaded all different types of videos, so people know we're normal. [P03]}\end{quote}


P03 posted videos of herself completing everyday tasks to demonstrate that she was "normal." Many participants, as prior research has shown \cite{liu_i_2019}, discussed the concept of being "ordinary." To blind individuals, being "normal" or "ordinary" represented their stance against being seen as a marginalized group in mainstream society. Moreover, the videos shared by BlindTokers that highlighted their abilities also included instructional information on how to navigate daily tasks as a blind person. Such instructional content was beneficial to their blind audience members who faced similar challenges. For instance, P39, a blind mother who overcame her fear of parenthood with the assistance of TikTok, stated:

\begin{quote}\textit{For a while, I was afraid to have kids. I would be afraid that I would lose them as we walk out of the house. But there's another mom on TikTok, she's been able to adapt to being parents... [I learned] some of the tips...like having your kids wear neon or bright colors... And I've found that so comforting that I could one day be a parent. [P39]}\end{quote}


In cases similar to P39, the videos created by blind parents that showcased their capabilities also helped BlindTokers learn how to be parents. Such content creation by BlindTokers, in addition to serving as instructions, also provided emotional support to their blind audience. For example, P20 shared her feelings about consuming videos of her friend (a female BlindToker):

\begin{quote}\textit{She goes out like she took her camera with her when she went on the plane, the Disneyland, going downstairs in her house. She does all kinds of different videos just to prove to people, "I can do this, I don't have to just sit in a bubble and be alone all day." I'm super proud of her! [P20]}\end{quote}


The vlog culture on TikTok encourages users, including BlindTokers, to share their daily lives. P20's friend was one of the BlindTokers who could both enjoy her life by trying various experiences and showcase these experiences to her audience. Witnessing the lives and adventures of other blind people on TikTok inspired many participants, like P20, which also contributed to BlindTokers' pride in their identity.

\subsection{Voicing Out for BlindTokers' Rights and Responsibilities}

In addition to posting content about their lived experiences, BlindTokers also shared their work in advocating for their rights and responsibilities. As a community consisting of minority individuals, BlindTokers often experienced marginalization, either in real life or on TikTok. Some of the stories in the last subsection also indicated this, such as encountering inaccessible building designs. Reflecting on this marginalization, BlindTokers decided to use TikTok as a platform to voice out for their minority identity. They collectively worked on discussing their experiences and seeking ways to improve their situations.


\subsubsection{\textbf{Revealing Violations of Blind People's Rights}} 


Numerous studies have pointed out that most of the modern world is inherently more suitable for sighted people \cite{natalie_uncovering_2021,jandrey_image_2021,tigwell_emoji_2020,gleason_making_2019}, and blind people's experiences are often overlooked. This manifests not only as accessibility issues but also as a lack of public awareness regarding such accessibility issues. Therefore, many participants (N=17) reported that they used TikTok to post content to increase the visibility of blind people. Specifically, the content aimed to raise awareness among their audience about the rights of blind individuals to enjoy an accessible world. For instance, some participants revealed the accessibility issues of TikTok, like P12 said:

\begin{quote}{\textit{(We) really can't line ourselves up in the video...So it doesn't really give us a cue or indication of how far we are, how close we are... I'm pretty sure there's been some type of viral video about TikTok not been accessible. [P12]}}\end{quote}


As mentioned in the previous section, BlindTokers enjoyed creating videos either for socializing or for entertainment. Such creation sometimes involves composition and the use of special filters. However, the accessibility issues in TikTok's technical designs made content creation challenging for BlindTokers. They also posted videos about these issues to raise awareness for accessible design and the accessibility of content created by sighted creators. P09 said:

\begin{quote}\textit{...some people have a video... `check out my dance on my new shirt,` but they did not say like `the shirt has a puppy on it and the puppy's drinking beer...` It can be a pain in the ass, I wish this community took more seriously. I know people wanna just immediately up and post something. But image description and video description help me know what's going on. [P09]}\end{quote}

P09 also pointed out that many TikTok creators ignored the accessibility of their content when making videos. This demonstrates a lack of attention to the blind community that needs to be addressed.




Other than drawing attention to accessibility issues on TikTok, some participants also used TikTok videos to address misinformation about blind people's legal rights regarding enjoying accessible outdoor experiences in real life. P15, a former lawyer, shared a story about the misunderstanding of blind people's legal rights:

\begin{quote}\textit{There is the belief that a person has to be declared legally blind by their state before they can use a white cane. [But the truth is] every state has a white cane law. It is about the fact that a person carrying a white cane has the right of way, no matter where they are crossing a street. [So I posted on TikTok to clarify]... it should just be labeled like either `the white cane` or `white cane law.` [P15]}\end{quote}


Misinformation on social media can be harmful. In this case, misinformation about the "white cane law" definitely harmed blind people's rights: it could affect blind people's intention to use white canes, and walking without white canes would increase the risk to their physical safety. P15 posted a video to clarify the misunderstanding using his own expertise. He was also aware of the algorithmic features of TikTok and used hashtags to reach a larger audience.

\subsubsection{\textbf{Deliberating Responsibilities for BlindTokers' Well-being}} 


Participants (N=12) also reported on how they deliberated their responsibilities as content creators for BlindTokers' well-being, such as ensuring accessible content on TikTok and displaying positive attitudes towards difficulties. Participants, being aware of the issues mentioned above, took it upon themselves to care for the community. Driven by such a strong sense of responsibility, they engaged in various efforts to improve the community and nurture an accessible and inclusive space for BlindTokers. P26 demonstrated how to be a more responsible creator for the BlindTok community:

\begin{quote}\textit{Most of the [my] videos don't have a ton of movement or change in scene. A lot of them are me just kind of sitting and talking to the camera. So there isn't too much of a visual aspect unless you want to see my face. Other than that it's all auditory. And I include captions for the people who have difficulty hearing or can't hear. [P26]}\end{quote}




As mentioned above, the inability to access content due to a lack of audio description can lead to disconnection and frustration. Recognizing these issues through personal experience, P26 reported being mindful while creating content on TikTok to ensure others would not face similar challenges. This practice is consistent with prior work \cite{bennett_its_2021}. Additionally, P26's efforts benefited not only the BlindTok community (by minimizing movement and changes of scenery) but also users with hearing impairments (by including captions). Many participants reported similar practices to make videos more accessible to people with hearing impairments (e.g., [P20]). This suggests that BlindTokers cared about not only people with visual impairments but also people with hearing impairments. Their identity work was inclusive and encompassed the overall community of people with disabilities rather than focusing solely on blind individuals.


In addition to enhancing the accessibility of TikTok videos, participants also demonstrated a sense of responsibility in helping blind users maintain a positive outlook. As reported by P06, "\textit{most of us, we don't really shout out about it (being blind), kind of keep it very much to ourselves}"; this suggests that blind people had experienced stigmatization in mainstream society and were hesitant to openly acknowledge or accept their identity. Therefore, some participants took it upon themselves to assist these blind users, such as P18, who hosted a show specifically designed to help blind individuals navigate challenging situations. He explained why and how he engaged in this effort:


\begin{quote}\textit{I like a lot of videos where people just come up on someone’s homeless and surprise them with money to just make their day better... I was like, `man, I gotta try to be like those people.` And that's another message of the [show name]... getting visually impaired, it’s not like a death sentence. You can still do these things, and we're gonna show you that you can. [P18]}\end{quote}

As we mentioned earlier, blind people can find inspiration in witnessing the lives of other BlindTokers. In P18's case, he regarded inspiring other BlindTokers as a personal responsibility. P18's story also highlighted the advantages of videos that conveyed positivity. A positive attitude can serve as an inspiration to others, and those who are inspired may, in turn, spread positivity to more people.

\subsection{Addressing Sighted People's Misconceptions}

Participants in this study also reported how they responded to misconceptions held by sighted people about BlindTokers. It's worth noting that in previous sections, we discussed how BlindTokers engaged in content creation to specify their lifestyles or clarify misunderstandings about blind people, which can be seen as efforts to address misconceptions. In this section, rather than detailing the specific content used for clarification, we will illustrate the interpersonal interactions of BlindTokers when they encounter misconceptions. These interactions will focus on the attitudes, strategies, and practices of BlindTokers in response to misconceptions. We will also highlight the positive emotions experienced by BlindTokers during their identity work, such as when they deal with trollers or educate friendly sighted individuals.

\subsubsection{\textbf{Fighting Against Ableism}} 


As mentioned earlier, many sighted people do not understand that blindness exists on a spectrum. According to numerous participants (N=19), this lack of understanding has resulted in ableist misconceptions. Ableism is defined as "\textit{a system that places value on people's bodies and minds based on societally constructed ideas of normalcy, intelligence, excellence, and productivity}" \cite{lewis_ableism_2020}. In our study, ableism manifested itself in critical attitudes and judgmental comments directed towards BlindTokers. Participants shared their experiences of encountering bias and discussed how they responded to it. One form of response involved documenting evidence of ableism and sharing it on TikTok to raise awareness, as explained by P09:

\begin{quote}\textit{The dude thought that I was faking (being blind)... because if I actually was blind I should not be using the phone, and then if I'm not blind I should not be using cane. That is not at all the first time that I've had those experiences. That's just the first time I was quick enough to tell Siri to start recording... And the video itself is like 10 minutes long. I edited it so that people could stitch or duet to do whatever to help [promote it]. [P09]}\end{quote}



P09 revealed that many ableists make assumptions about the lifestyles of people with disabilities, and individuals with disabilities often face criticism if they do not conform to these assumptions. This conflict can result in practices or behaviors that harass blind people. P09's response to such harassment emphasized the importance of documenting evidence and sharing it to raise awareness of ableism. He also demonstrated his strategies for leveraging TikTok's features to facilitate collaboration in spreading this evidence. In addition to recording and sharing, some participants mentioned that they offered support to their blind friends when they were attacked. P20 recalled an incident where her friend, who was also a BlindToker, faced harassment from sighted people on TikTok:

\begin{quote}\textit{(There's a) complete jerk saying (to my friend) "you're faking, you're lying, you can do this, you are not blind"... I tell them off in a very, not nice way. I do not want to hide my feelings...I just comment underneath. And when I type wrong text, then someone will say `oh my god, you use the wrong word there` and I'm like `way to go, make fun of the blind person for talk texting (typing text for conversation).` [P20]}\end{quote}

As mentioned in previous sections, P20 witnessed and was inspired by her friend, another BlindToker's self-presentation. She demonstrated consideration for her friends with hearing impairments by adding captions to her videos. These experiences indicate that P20 placed value on her friendships with TikTokers with disabilities. Therefore, her act of assisting her friend who was facing harassment also reflects her care for them. Furthermore, the experiences related to incorrect text in P20's comments highlighted TikTok's inaccessible commenting functions. Additionally, some participants discussed how they leveraged TikTok's design to collectively combat ableism at the community level. P08 shared his approach:

\begin{quote}\textit{If there's another person who is accusing us faking our disability. I will stitch that video and call people out...I will be showing support by liking and replying to them. And if there is a comment that I have issues with, you can reply to that comment with the video, so that people can see the comment. [P08]}\end{quote}


Similar to P20, P08 also exhibited his support for BlindTokers who faced harassment. His efforts extended beyond caring for friends and emphasized the solidarity of the entire community. To involve more people in defending the community, he utilized TikTok's algorithmic features. Comments and the number of "likes" were seen as expressions of support for BlindTokers, while "stitches" and "reply with videos" allowed for the location of videos or comments, making it easier for people to identify ableist content and individuals more easily.


\subsubsection{\textbf{Clarifying Sighted People's Genuine Confusions}} 



Participants (N=35) also mentioned responding to friendly questions from sighted people. While participants encountered many ableists, a significant portion of their audience consisted of friendly sighted individuals who had little knowledge about blind people. These individuals would leave comments on BlindTokers' videos or create their own videos to ask questions. Participants were willing to answer these questions. However, due to the context of social interactions, the intention behind some questions was not always clear. Sometimes the questions were genuinely curious, while other times they might have been implicit trolling. For instance, P06 recalled someone asking questions about blind people's toileting habits. Although it was a question related to blind people's lifestyles, it still made P06 feel insulted because it was private and suggestive. Therefore, BlindTokers also undertook additional work to discern which questions were friendly and which were not. P06 elaborated on how she identified the tone of these questions:

\begin{quote}\textit{It's difficult to tell when people type things... [whether] they were trolling or just being silly...you don't always get the right tone. But when you use voice... you can imply what they're saying. And TikTok is primarily videos, it's much easier to identify people's tones. [P06]}\end{quote}


P06 relied on verbal cues to assess the intent behind questions from sighted users. TikTok played a helpful role for blind users in this regard. Once blind users discerned that the questions were genuine and not trolling, they would respond sincerely. P42, who enjoyed sharing aspects of blind people's lives, described how she interacted with these questions:

\begin{quote}\textit{\textbf{P42}: So being able to tell people on the Internet the real ins and outs of it (lives with service dogs)...being able to educate somebody on something so small can really make such a difference... How many sighted people [did I answer]? I had no idea.}\end{quote}
\begin{quote}\textit{\textbf{Interviewer}: Does TikTok make answering the questions easier?}\end{quote}
\begin{quote}\textit{\textbf{P42}: 100\%, because typing for me takes forever. So if I can just spill some words out in a minute (reply with video), instead of having to sit there for five and type out what I wanted to say, absolutely, it does make it easier.}\end{quote}


P42 answered numerous questions from sighted people on TikTok, and through these interactions, she also dispelled misunderstandings about her identity. She, along with many other participants, referred to these Q\&A exchanges, which contributed to sighted people's improved understanding of blind individuals, as "education." This indicates that BlindTokers derived a strong sense of satisfaction from engaging in these friendly and informative interactions with sighted individuals. Additionally, P42 highlighted that TikTok's features facilitated this "educational" work.




\section{Discussion}


Next, we will discuss the findings within the context of the existing literature from two major perspectives. Firstly, we identify a salient pattern in the participants' desires to present, inspire, and echo BlindTokers' lives with strong positivity. We have observed that this pattern aligns with the notion of flourishing, as outlined in the positive design framework. We will then delve into how BlindTokers conceptualize the \textit{state of flourishing}, especially as a community on TikTok. 

\begin{figure*}[htp]
    \centering
    \includegraphics[width=\textwidth]{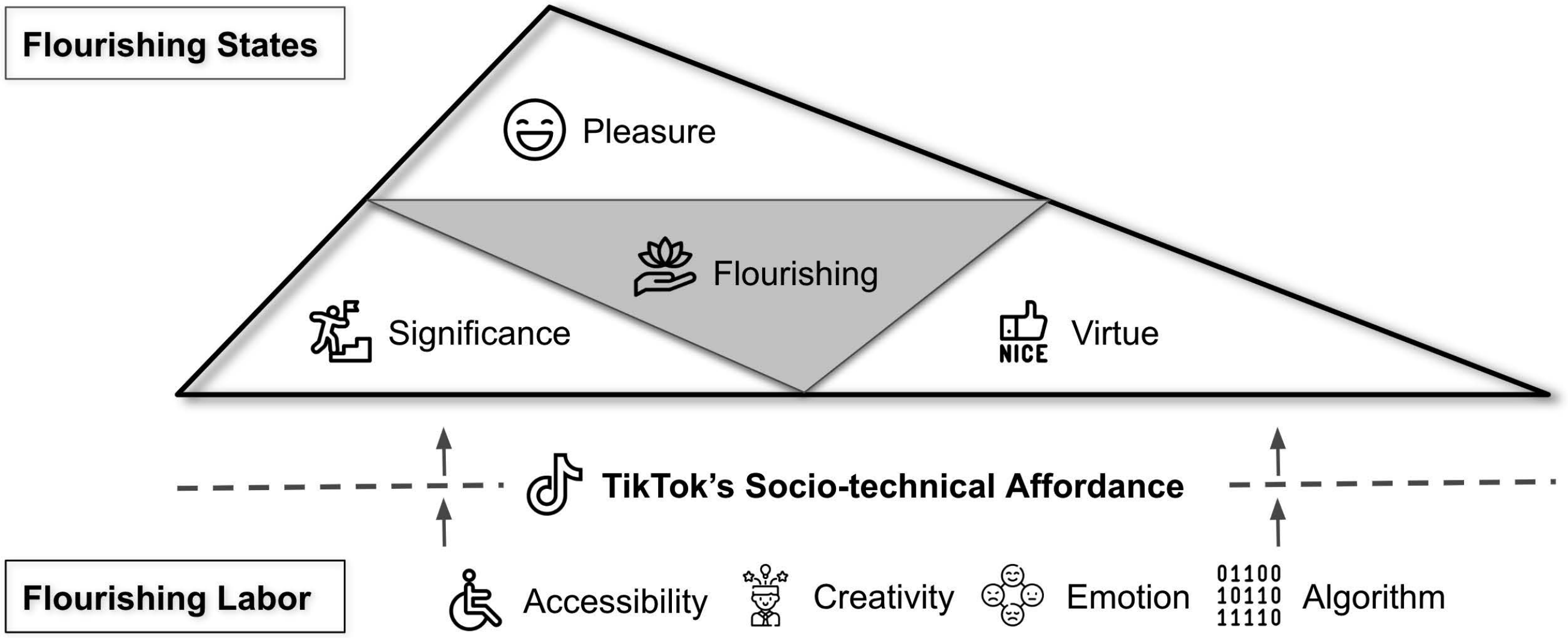}
    \caption{Flourishing States and Flourishing Labor}
    \label{discussion}
    \Description{Flourishing States and Flourishing Labor}
\end{figure*}

To be noted, using a flourishing perspective does not mean we are only interested in the positive side of BlindTokers' identity; rather, we also pay attention to the negative parts that hinder BlindTokers from flourishing. We have also observed that to attain a state of flourishing, BlindTokers must invest tremendous efforts in terms of developing strategies and engaging in practices. Therefore, we will also adopt a \textit{labor} lens to highlight the efforts required for flourishing, which we will refer to as \textit{flourishing labor}.(Figure \ref{discussion})

\subsection{Flourishing of BlindTok: The Desired States of BlindTokers on Their Collective Identity}



From the findings, we have identified a significant pattern in the motivations behind BlindTokers' identity work. As members of the BlindTok community, they aspire to portray their lives with a strong emphasis on positivity, aiming to inspire and resonate with the positive aspects of others' lives. These activities not only highlight the challenges of an inaccessible world but also provide a source of enjoyment and a profound sense of fulfillment for BlindTokers. To analyze BlindTokers' desire for their collective identity, we draw upon the notion of \textit{flourishing}. As described by Desmet and Pohlmeyer \cite{desmet_positive_2013}, flourishing encompasses the intersection of design for virtue, pleasure, and personal significance. By illustrating the three states of flourishing, we will also discuss TikTok design that mediated the flourishing.

1) \textbf{Pleasure}: Pleasure involves the increase of positive sentiments and/or the decrease of negative sentiments \cite{johnson_joy_2020}. It usually represents one's temporary emotions \cite{desmet_positive_2013}. In our findings, BlindTokers' pursuit of pleasure included creating and sharing jokes related to blind individuals. They also took pleasure in exploring each other's personalities beyond blindness. TikTok's playful features, such as voice-over acting, special audio effects, and duets, allowed BlindTokers to enjoy such feelings; and the comments and live-streams enabled them to share these feelings. This is consistent with a recent study, where Schaadhardt et al. \cite{schaadhardt_laughing_2023} demonstrated that young people on TikTok used humor to address their mental health issues and build communities. However, several participants voiced concerns that videos without audio descriptions were inaccessible, highlighting the importance of accessibility in humor, especially when presented on social media platforms.

2) \textbf{Personal Significance}: Personal significance can be achieved through a sense of progress that propels individuals toward their desired accomplishments \cite{desmet_positive_2013}. As we presented in the findings, BlindTokers held various expectations, either for themselves or the community. As individuals, they aspired to achieve independence in parenting, while as members of the BlindTok community, they aimed to enhance the experiences of other BlindTokers on TikTok and in real life, to educate the sighted community to foster better understanding of the blind community, and to combat ableism with solidarity. They experienced a sense of personal significance when they fulfilled these expectations. TikTok's design elements, such as hashtags, filters, video sharing, "like" buttons, and "reply with videos" functionality, facilitated the processes mentioned above.

3) \textbf{Virtue}: Virtue emphasizes happiness resulting from one's moral and ethical behaviors \cite{desmet_positive_2013}. BlindTokers held moral requirements, especially within the TikTok platform. They made sense of the socio-technical environment of TikTok by observing other TikTokers' activities and engaging with the audience. Combining their real-life experiences with their TikTok experiences, they developed a set of moral obligations and adhered to them. They admired individuals who demonstrated kindness, positivity, and inspiration to others, which influenced them to create similar content. Furthermore, they encountered accessibility issues on TikTok, leading them to consider it a moral obligation to produce content that is accessible. Notably, BlindTokers' moral obligations regarding accessibility extended to caring for not only blind individuals but also those with other disabilities. TikTok's technical features, such as following, sharing, stitching, and auto-captioning \cite{10.1145/3579490, mcdonnell_understanding_2022, macleod_understanding_2017}, played a significant role in supporting these moral endeavors.

TikTok played a dual role in both empowering and shaping the flourishing of the BlindTok community. On one hand, TikTok's user base and its design for content creation and sharing facilitated the connection of BlindTokers with a large community of blind users. This collective engagement resulted in the formation of the BlindTok community, where members showcased their lives, which were observed by others on TikTok. Within this community, expectations and desires for flourishing began to take shape. However, it's important to note that these desires for flourishing, as reported by participants, were also influenced and framed by the socio-technical limitations of TikTok. The accessibility challenges within TikTok, as well as interactions with sighted users, served as primary factors that informed BlindTokers' expectations for the states of flourishing.

That said, it's worth noting that while some flourishing experiences can be easily attained (e.g., giving a speech or responding to others' comments with a "talking head" style video), many of these experiences demanded substantial effort. This effort often involved using various TikTok design elements to create content that catered to blind users. In the following section, we will emphasize the labor and effort that underpinned the flourishing of BlindTok and highlight design directions that can facilitate this work.







\subsection{Flourishing Labor: BlindTokers' Work to Support Flourishing States}




As previously illustrated, BlindTokers harbored diverse desires concerning their identity, with a primary focus on fostering the flourishing of their community. However, as our findings indicate, realizing these desires was far from easy, and it became even more demanding when a large community sought to flourish together. Especially, some BlindTokers' work went unnoticed by BlindTokers or lacked support from the platform. For example, while all BlindTokers wanted accessible content, few of them (12 out of 45) mentioned consciously making their content accessible to other blind users. In this study, we adopt a \textit{labor} perspective to foreground the work undertaken by BlindTokers in cultivating and sustaining flourishing states. The labor perspective places emphasis on the contributions made by individuals, raises awareness of collaboration for collective efforts, offers practical guidance for collaboration, and establishes a sustainable workflow within the community \cite{simpson_rethinking_2023, steiger_psychological_2021,rosenblat_algorithmic_2016,psarras_its_2022,dosono_moderation_2019}. We introduce the concept of "\textit{flourishing labor}" to describe the efforts made by BlindTokers that contribute to the flourishing states of the BlindTok community, particularly in helping them realize their desires for pleasure, personal significance, and virtue. We also propose design implications to facilitate and support these efforts.

1) \textbf{Flourishing Labor for Accessible Content Creation and Consumption}. The most crucial aspect of flourishing labor lies in the efforts to ensure access, which involves making TikTok and its content accessible to BlindTokers \cite{alharbi_accessibility_2023}. In recent years, scholars in accessible computing have increasingly emphasized building interdependence among individuals with varying disability statuses rather than focusing solely on the independence of people with disabilities. This perspective encourages viewing the entire population as a collective where all members mutually support one another \cite{Bennett2018}. Our findings align with this perspective, as some BlindTokers intentionally created accessible content to accommodate individuals with different disabilities. However, our research also suggests that not all creators among BlindTokers were aware of the importance of accessibility labor. This highlights the need for future design efforts that promote awareness of accessibility and provide support for the creation of accessible content.

\textbf{Design Implications.} The accessibility design should focus on both BlindTokers' content creation and consumption. TikTok can apply different types of reminders, including visual or verbal ones, during BlindTokers' entire content creation process. For the content that has not been posted yet, TikTok should remind the creators of the accessibility considerations before they post it (adding captions or audio descriptions before posting the content); for the content already posted, TikTok should allow editions that make the content accessible (adding captions or audio descriptions after post the content. 


2) \textbf{Flourishing Labor for Creative Expression}. BlindTokers were motivated to create imaginative content tailored to entertain blind audiences, and this played a vital role in their flourishing as it contributed to the pleasure of blind viewers. This content could take various forms, such as humor that resonated with BlindTokers or a vivid portrayal of their visual impairment. This demonstrates the creators' profound understanding of both blind users and TikTok's creative tools. Furthermore, a recent study on creative labor within TikTok has highlighted that the platform's high demand for creative content could potentially deter individual creators from pursuing creative content creation with the right intentions \cite{simpson_rethinking_2023}. This underscores the need for a more distributed architecture of creative content creation, implying that we should not rely solely on a small group of creators to continually generate engaging content for the community.

\textbf{Design Implications.} To reduce the labor in creative expression, we propose designs that are both accessible and entertaining for BlindTokers. TikTok could incorporate a range of audio memes (interesting and trending sound effects) into the content creation interface, similar to the visual effect filters. This would provide BlindTokers with a rich palette of audio options to enhance their content. Additionally, designers might consider the integration of virtual avatars to support creative self-presentation of identity. To address the issue of over-reliance on a small group of creators, TikTok could actively encourage a more diverse pool of creators. TikTok could also recognize contributors by inviting them to share motivations, strategies, and techniques for producing creative content. Such events could be facilitated through TikTok functions, e.g., live-streaming, to foster knowledge sharing and collaboration among creators.

3) \textbf{Flourishing Labor for Emotional Support.} Some BlindTokers also worked on managing their emotions. They consistently exhibited positive attitudes to inspire other BlindTokers, and notably, they maintained a friendly demeanor when interacting with individuals who posed questions, especially sighted individuals who had limited knowledge about blind people. Many BlindTokers considered it their duty to raise awareness about the BlindTok community, which often led them to encounter numerous questions from sighted individuals. While it was justifiable for BlindTokers to express frustration when dealing with ableists who posed humiliating questions, they also extended kindness to those who genuinely sought to learn. In such interactions, they had to discern the intentions of the person involved. They strived to remain polite and friendly while distinguishing between whether someone was an ableist or a friendly individual. As per the accounts of some BlindTokers, TikTok facilitated this process because most of the content was in video format, allowing BlindTokers to easily discern people's tones and intentions through visual cues. However, they faced a challenge with text comments, as it was often difficult to discern the intentions behind written messages.


\textbf{Design Implications.} To better support the identification of tone in non-verbal content, such as comments, we suggest providing pre-designed comment templates that instruct the audience to craft straightforward and friendly questions for BlindTokers. We also recommend that TikTok implement crowd-sourcing mechanisms like "down-votes" \cite{lampe_slashdot_2004} to allow other TikTokers to collectively help identify unfriendly tones in comments. Comments with a significant number of down-votes should be hidden or displayed with a mark that reminds BlindTokers that these comments might involve harassment. Additionally, the platform can leverage its powerful algorithm to include educational videos in its video recommendation feed. These videos can raise public awareness about ableism, allowing users to periodically watch them to promote understanding and empathy, thereby reducing harassment and emotional labor.



4) \textbf{Flourishing Labor for Algorithmic (In)visibility}. The last point addresses the algorithmic aspect of flourishing labor on TikTok. Based on our findings, BlindTokers adopted various practices to make the community algorithmically visible \cite{abidin_mapping_2020}. They promoted other's content by liking, commenting, and sharing. They promoted their content using hashtags and editing video lengths. However, it's worth noting that while working on improving algorithmic visibility, BlindTokers may also attract unwanted attention from trolls. In a recent study on disability activism on social media \cite{sannon_disability_2023}, researchers expressed concerns about the risks faced by people with disabilities when voicing out on social media. This highlights the importance of managing algorithmic visibility, particularly for those who are influenced by TikTok. Some participants have shared techniques for decreasing algorithmic visibility to trolls. These techniques include blocking, reporting, and controlling their audience. However, these functions also had accessibility issues.

\textbf{Design Implications.} Previous research has demonstrated that TikTok intentionally manipulates its algorithm to influence the visibility of specific groups on its platform \cite{zeng_content_2022}. Strategies such as de-platforming or de-recommending \cite{ma_defaulting_2023} have the effect of preventing new users from discovering and joining these communities. To empower BlindTokers for better control over the algorithmic visibility of their community, we suggest that TikTok promote more accessible content explaining the inner workings of TikTok's algorithm. This would provide BlindTokers with a better understanding of how to navigate the algorithmic environment, whether they wish to be more visible or less so. Consequently, it is essential to offer more accessible features that allow users to manage their algorithmic visibility effectively.

\section{Limitations and Future Work}
The primary limitation of this study is about sampling methods. We collected data from 45 blind users who are mostly from the USA. Considering that we are studying the identity presentation of Blind Tokers, cultural influence could be an important factor that shapes participants' perceptions of identities. For instance, while experiencing inaccessible designs or harassment from sighted users, our participants chose to voice out their experiences and confront ableists. However, as reported by another study that investigated users on Douyin \cite{rong_it_2022}, the Chinese TikTok, users preferred passive strategies like hiding their identities while encountering harassment. Therefore, different cultures might impact the work and attitudes toward identity. Future studies on blind users' identity work can pay more attention to cultures outside the USA to address the limitations of the current study.
\section{Conclusion}
People utilize social media as an indispensable means to form online communities and present their shared identities. HCI and CSCW scholars have been showing increasing interest in the presentation of the identities of people who are marginalized due to racism or sexism. But little is known about the identity work of people with disabilities. In this paper, we present an interview study on 45 blind users on TikTok. We find that blind users carry out three types of identity work, including demonstrating characteristics, voicing out their rights and responsibilities, and clarifying sighted people's confusions. We also discuss the findings with the literature on flourishing and provide design implications for flourishing labor.

\begin{acks}
Thanks for the reviewers' comments.
\end{acks}

\bibliographystyle{ACM-Reference-Format}
\bibliography{Yao}

\end{document}